# Nonlinear localized excitations in a topological ferromagnetic honeycomb lattice

Wenhui Feng[1], Bing Tang[1,2]* , Lanjun Wu[1], Peng Kong[1], Chao Yang[1],

Lei Wang[3], Ke Deng[1]

[1]Department of Physics, Jishou University, Jishou 416000, People' Republic of China

[2]The Collaborative Innovation Center of Manganese-Zinc-Vanadium Industrial Technology, Jishou University, Jishou 416000, People' Republic of China

[3] Department of Mathematics and Physics, North China Electric Power University, Beijing 102206, People' Republic of China

**ABSTRACT**

We theoretically investigate nonlinear localized excitations in a Heisenberg honeycomb ferromagnet with a second neighbour Dzialozinskii-Moriya interaction, which has been proved to possess the topological band structure. Using the time-dependent variational principle, we obtain the equation of motion for the system in the Glauber coherent-state representation. By means of the semidiscrete multiple-scale method, different types of nonlinear localized excitations are gotten. Our results show that both two-dimensional discrete breathers and bulk gap solitons may occur in the Heisenberg honeycomb ferromagnet.

## 1. Introduction

As is well known, the graphene is a two-dimensional (2D) single-layer periodic carbon material with a honeycomb lattice structure [1]. One of the main properties of a monolayer graphene is that two disparate energy bands can get close to each other at

* Corresponding author.
E-mail addresses: bingtangphy@jsu.edu.cn;bingtangphy@163.com

Dirac points. The honeycomb crystal structure of the graphene can cause some very important phenomena about the electronic transport properties of this 2D material. Especially, Kane and Mele [2] have found that the monolayer graphene with spin-orbit coupling can reveal a nontrivial topological band structure containing an energy gap at the Dirac points, which can lead to the quantum Hall effect linked to topologically protected edge electron states. In general, the idea of topological energy band structure does not depend on the statistical property of the quasiparticles so that it can be expanded to bosonic systems such as photons[3], phonons [4], and magnons [5].

In the past few years, the topological nature of magnons in ordered magnets has attracted more and more attention from physicists and mathematicians. The Heisenberg ferromagnet on the 2D honeycomb lattice is of particular interest since its magnon bulk bands can reveal six Dirac points at the corners of the first Brillouin zone[6]. Furthermore, it has been shown that the corresponding Bogoliubov Hamiltonian in the vicinity of the Dirac points can realize a massless 2D Dirac-like Hamiltonian, which possesses a time-reversal symmetry. Owerre [7] has found that a gap can form at the Dirac points by introducing a second neighbour Dzyaloshinskii–Moriya (DM) interaction, which destroys the inversion symmetry of the honeycomb lattice and the time-reversal symmetry of the Bogoliubov Hamiltonian. By this simply way, he has theoretically realized the honeycomb topological magnon insulator with nontrivial edge states for the first time. Inspired by Owerre's work, Pantaleón and Xian have analytically investigated the properties of magnon edge states in a semi-infinite ferromagnetic honeycomb lattice with the second neighbour DM interaction[8-10]. In their works, zigzag boundary, bearded boundary, and armchair boundary have been considered, respectively. Their results have showed that the energy spectra of the magnon edge states become tunable when an on-site anisotropy of the outermost sites is introduced. In fact, except for the DM interaction, the spin-anisotropic interaction (i.e., the Kitaev interaction[11]) can also open a gap between two magnon bands for the honeycomb ferromagnet. Particularly, several researchers have displayed that the ferromagnetic Kitaev-Heisenberg model (without the DM interaction) on the honeycomb lattice can also support topologically protected magnon edge states[12,13].

It is important to mention that, in honeycomb ferromagnet, both the spin-spin exchange interaction and the DM interaction are intrinsically nonlinear. However, a majority of works on topological excitations theoretical works have been only focused on the linear region. Until very recently, Elyasi *et al.* [14] have numerically studied the existence and properties of self-consistent solitons in a ferromagnetic Heisenberg honeycomb lattice with the topologically nontrivial band gap. Some new results have been obtained. For example, by calculating the bulk soliton generation phase diagram, they have suggested that phase boundaries between bulk solitons with trivial and nontrivial topological properties may exist. Their work opens a new research direction, namely, nonlinear localized excitations in honeycomb ferromagnets with the topological band structure.

In our work, we will present an analytical study on nonlinear excitations in a topological ferromagnetic Heisenberg honeycomb lattice with a second neighbour Dzialozinskii-Moriya interaction in the semiclassical limit. Our aim is to seek for analytical forms of nonlinear localized excitations in the Heisenberg honeycomb ferromagnet and ascertain their existence conditions and characteristics. In Section 2, starting from the Hamiltonian of the ferromagnetic Heisenberg honeycomb spin lattice mode, the bosonized form for the honeycomb ferromagnetic spin lattice model Hamiltonian shall be gotten via taking advantage of the Dyson-Maleev transformation. In Glauber's coherent-state representation, the equation of motion for the coherent amplitude is derived by the use of the time-dependent variational principle. In Section 3, the envelope function of the coherent state amplitude is obtained by employing the semidiscrete multiple-scale method. In Section 4, both the Brillouin zone center modes and nonlinear excitations near the Dirac points are analytically studied. The whole work is summarized in the last section.

**2. The ferromagnetic honeycomb lattice model and equation of motion**

In this work, let us consider the following Hamiltonian for the honeycomb ferromagnet

$$H = H_{\mathrm{H}} + H_{\mathrm{DM}} + H_{\mathrm{ext}}. \tag{1}$$

First, we write down the first term in Eq.(1), which reads

$$H_{\mathrm{H}} = -J \sum_{\langle i,j \rangle} \mathbf{S}_i \cdot \mathbf{S}_j. \tag{2}$$

This term corresponds to the nearest neighbor Heisenberg spin-spin exchange interactions.

The second contribution in Eq.(1) describes the DM interaction [16,17] between next-nearest neighbors, which is antisymmetric. The corresponding Hamiltonian has the following form

$$H_{DM} = D \sum_{\langle\langle i,j \rangle\rangle} v_{ij} \mathbf{e}_z \cdot \left( \mathbf{S}_i \times \mathbf{S}_j \right), \tag{3}$$

where $v_{ij} = \pm 1$ is an orientation-dependent parameter in analogy with the Kane–Mele model [2]. Actually, the next-nearest neighbor DM interaction is equivalent to the spin-orbit coupling, which can destroy the inversion symmetry of the honeycomb lattice [18].

Here, we introduce the coupling to an external magnetic field aligned with the z axis via a Zeeman term, namely,

$$H_{ext} = -g\mu_B \sum_i H_z S_i^z. \tag{4}$$

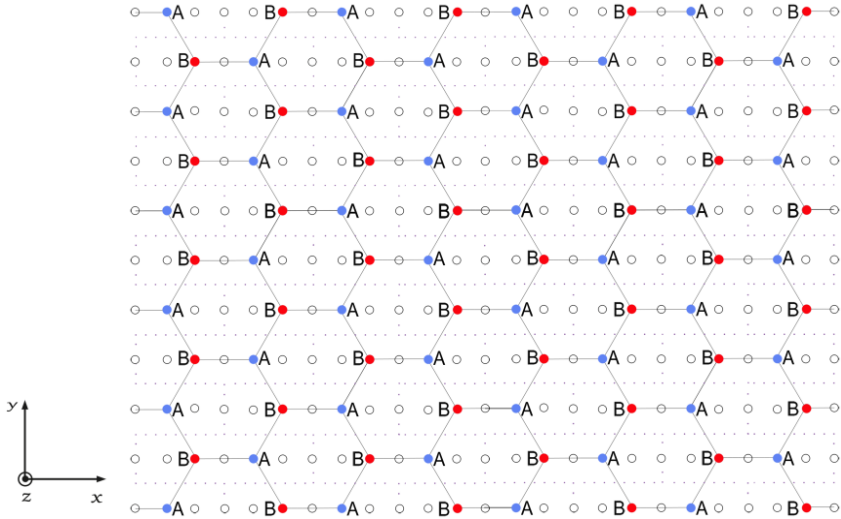

**Fig. 1.** Illustration of the honeycomb lattice structure on the x–y plane. Color solid circles stand for the sites in the honeycomb lattice, black open circles indicate the unused sites in the underlying rectangular lattice. The dotted lines represent the unit cells, each of which has two nonequivalent sites A and B.

Our study shall be focused on a two-dimensional honeycomb lattice, which lies on a subset of one underlying rectangular lattice. Referring to Rf. [15], we first need to introduce the orthonormal basis vectors $\mathbf{e}_x$ and $\mathbf{e}_y$ so as to describe the position of the $(m,n)$ site for the rectangular lattice. Thus, the corresponding position vector can be written as $m\mathbf{e}_x + hn\mathbf{e}_y$ with $h=\sqrt{3}$. For the sake of specifying the two-dimensional honeycomb lattice, we should reserve only those sites $(m,n)$ that satisfy $m+n = even$ and abandon $(m=6p+1, n=odd)$ and $(n=6p+4, n=even)$. In Fig. 1, the color solid circles represent the sites reserved in the two-dimensional honeycomb lattice that follow these relations, and the black open circles exhibit all the rest of sites in the rectangular grid. In this work, we focus on a Heisenberg ferromagnet lying on this honeycomb lattice, which includes the nearest neighbor Heisenberg spin-spin and the next nearest neighbor DM interactions. In principle, the honeycomb lattice has C$_3$ rotational symmetry, being composed of tessellating hexagons, where rotation by any angle of a multiple of $2\pi/3$ leaves the lattice invariant. For the present honeycomb ferromagnetic lattice, its inversion symmetry is destroyed due to introduction of the DM interaction. Note that the present honeycomb ferromagnetic lattice contains two different arrangements consisting of three nearest neighbour sites, as displayed in Fig. 2. From this figure, we can clearly see that each site is linked to three sites in different arrangements.

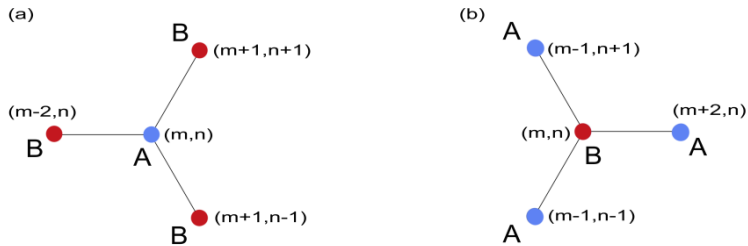

**Fig. 2.** Labelling of the sites in the honeycomb lattice: (a) Arrangement 1, $\mathbf{S}^A_{m,n}$ in centre, three neighbouring sites are $\mathbf{S}^B_{m-2,n}$, $\mathbf{S}^B_{m+1,n-1}$, and $\mathbf{S}^B_{m+1,n+1}$; (b) Arrangement 2, $\mathbf{S}^B_{m,n}$ in centre, three neighbouring sites are $\mathbf{S}^A_{m+2,n}$, $\mathbf{S}^A_{m-1,n-1}$, and $\mathbf{S}^A_{m-1,n+1}$.

Considering that each unit cell has two spins $\mathbf{S}^A$ and $\mathbf{S}^B$, we introduce the two sets of magnon creation (annihilation) operators $a^+(a)$ and $b^+(b)$, which are corresponding to the A and B sublattices, respectively. To study the nonlinear effect of the present honeycomb ferromagnet, we perform the Dyson-Maleev transformation[19,20] for the two sublattices A and B:

$$S^{A+} = S^{Ax} + iS^{Ay} = \sqrt{2S}(1 - \frac{a^+ a}{4S})a, \tag{5a}$$

$$S^{A-} = S^{Ax} - iS^{Ay} = \sqrt{2S}a^+, \tag{5b}$$

$$S^{Az} = S - a^+ a \tag{5c}$$

(similarly for B spins). Here, the Glauber coherent-state [21] shall be chosen as a basic representation so as to describe the components of the state vector of the magnon system. In the Glauber coherent-state representation, the state vector $|\Psi(t)\rangle$ can be written as follows:

$$|\Psi(t)\rangle = \left(\prod_{\text{A-sublattice}} |\alpha_{m,n}\rangle\right)\left(\prod_{\text{B-sublattice}} |\beta_{m,n}\rangle\right), \tag{6}$$

where coherent-state amplitudes $\alpha$ and $\beta$ satisfy $a|\Psi(t)\rangle = \alpha|\Psi(t)\rangle$ and $b|\Psi(t)\rangle = \beta|\Psi(t)\rangle$. By means of the time-dependent variational principle[22], the equations of motion for coherent-state amplitudes can be derived. For A sites in arrangement 1, one has

$$i\frac{d\alpha_{m,n}}{dt} = -JS(\beta_{m-2,n} + \beta_{m+1,n+1} + \beta_{m+1,n-1}) + \omega_0 \alpha_{m,n}$$
$$+ iDS(\alpha_{m+3,n+1} - \alpha_{m+3,n-1} - \alpha_{m,n+2} + \alpha_{m,n-2} + \alpha_{m-3,n+1} - \alpha_{m-3,n-1})$$
$$+ J\frac{\beta_{m-2,n}^* \alpha_{m,n}^2}{4} + J\frac{|\beta_{m-2,n}|^2 \beta_{m-2,n}}{4} - J\alpha_{m,n}|\beta_{m-2,n}|^2 + J\frac{\beta_{m+1,n+1}^* \alpha_{m,n}^2}{4}$$
$$+ J\frac{|\beta_{m+1,n+1}|^2 \beta_{m+1,n+1}}{4} - J\alpha_{m,n}|\beta_{m+1,n+1}|^2 + J\frac{\beta_{m+1,n-1}^* \alpha_{m,n}^2}{4} + J\frac{|\beta_{m+1,n-1}|^2 \beta_{m+1,n-1}}{4} - J\alpha_{m,n}|\beta_{m+1,n-1}|^2$$
$$- iD\frac{|\alpha_{m+3,n+1}|^2 \alpha_{m+3,n+1}}{4} + iDS\frac{\alpha_{m+3,n+1}^* \alpha_{m,n}^2}{4} + iD\frac{|\alpha_{m+3,n-1}|^2 \alpha_{m+3,n-1}}{4} - iD\frac{\alpha_{m+3,n-1}^* \alpha_{m,n}^2}{4}$$
$$+ iD\frac{\alpha_{m,n+2}^* \alpha_{m,n+2}^2}{4} - iD\frac{\alpha_{m,n+2}^* \alpha_{m,n}^2}{4} - iD\frac{|\alpha_{m,n-2}|^2 \alpha_{m,n-2}}{4} + iD\frac{\alpha_{m,n-2}^* \alpha_{m,n}^2}{4}$$
$$- iD\frac{|\alpha_{m-3,n+1}|^2 \alpha_{m-3,n+1}}{4} + iD\frac{\alpha_{m-3,n+1}^* \alpha_{m,n}^2}{4} + iD\frac{|\alpha_{m-3,n-1}|^2 \alpha_{m-3,n-1}}{4} - iD\frac{\alpha_{m-3,n-1}^* \alpha_{m,n}^2}{4},$$

(7)

where $\omega_0 = 3JS + g\mu_B H_{\text{ext},z}$. What is more, the equation of motion corresponding to the B sites in arrangement 2 has the following form

$$i\frac{d\beta_{m,n}}{dt} = -JS(\alpha_{m+2,n} + \alpha_{m-1,n+1} + \alpha_{m-1,n-1}) + \omega_0 \beta_{m,n}$$
$$- iDS(\beta_{m+3,n+1} - \beta_{m+3,n-1} - \beta_{m,n+2} + \beta_{m,n-2} + \beta_{m-3,n+1} - \beta_{m-3,n-1})$$
$$+ J\frac{\alpha_{m+2,n}^* \alpha_{m+2,n}^2}{4} + J\frac{\alpha_{m+2,n}^* \beta_{m,n}^2}{4} - J|\alpha_{m+2,n}|^2 \beta_{m,n} + J\frac{|\alpha_{m-1,n+1}|^2 \alpha_{m-1,n+1}}{4} + J\frac{\alpha_{m-1,n+1}^* \beta_{m,n}^2}{4}$$
$$- J|\alpha_{m-1,n+1}|^2 \beta_{m,n} + J\frac{|\alpha_{m-1,n-1}|^2 \alpha_{m-1,n-1}}{4} + J\frac{\alpha_{m-1,n-1}^* \beta_{m,n}^2}{4} - J|\alpha_{m-1,n-1}|^2 \beta_{m,n}$$
$$+ iDS\frac{|\beta_{m+3,n+1}|^2 \beta_{m+3,n+1}}{4S} - iDS\frac{\beta_{m+3,n+1}^* \beta_{m,n}^2}{4S} - iDS\frac{|\beta_{m+3,n-1}|^2 \beta_{m+3,n-1}}{4S} + iDS\frac{\beta_{m+3,n-1}^* \beta_{m,n}^2}{4S}$$
$$+ iDS\frac{|\beta_{m-3,n+1}|^2 \beta_{m-3,n+1}}{4S} - iDS\frac{\beta_{m-3,n+1}^* \beta_{m,n}^2}{4S} - iDS\frac{|\beta_{m-3,n-1}|^2 \beta_{m-3,n-1}}{4S} + iDS\frac{\beta_{m-3,n-1}^* \beta_{m,n}^2}{4S}$$
$$- iDS\frac{|\beta_{m,n+2}|^2 \beta_{m,n+2}}{4S} + iDS\frac{\beta_{m,n+2}^* \beta_{m,n}^2}{4S} + iDS\frac{|\beta_{m,n-2}|^2 \beta_{m,n-2}}{4S} - iDS\frac{\beta_{m,n-2}^* \beta_{m,n}^2}{4S}.$$

(8)

It should be mentioned that both the discreteness and nonlinearity of the present honeycomb ferromagnet have been completely included in Eqs. (7) and (8).

## 3. Asymptotic analysis and nonlinear amplitude equation

Since in general it is impossible to get the exact solution to nonlinear lattice equations of motion like Eqs. (7) and (8), we shall look for an approximate analytical solution to the Eqs. (7) and (8) by means of the semidiscrete multiple-scale method. In 1986, Remoissenet [23] proposed this method to investigate soliton solutions in one dimensional nonlinear lattices. Later, it has been extended by Wattis to the higher dimensional nonlinear lattices[15,24-27]. According to Wattis's procedure, the first step is to rescale the current variables $m$, $n$ and $t$ by introducing the following variables

$$x = \rho m, \quad y = \rho h n, \quad \tau = \rho t, \quad T = \rho^2 t, \tag{9}$$

where $\rho \ll 1$ stands for the relative amplitude of the nonlinear excitation, and the spatial variables $x$, $y$ can be viewed as continuous variables in the following text.

Mathematically, different ansatzes for the A and B sites, hence we analyze each sublattice individually with Eqs. (7) and (8). For A sites, we look for solutions of the form

$$\alpha_{m,n}(t) = \rho e^{i\phi} f(x,y,\tau,T) + \rho^2 [g_0(x,y,\tau,T) + e^{i\phi} g_1(x,y,\tau,T) + e^{2i\phi} g_2(x,y,\tau,T)]$$
$$+ \rho^3 [h_0(x,y,\tau,T) + e^{i\phi} h_1(x,y,\tau,T) + e^{2i\phi} h_2(x,y,\tau,T) + e^{3i\phi} h_3(x,y,\tau,T)] + \ldots,$$

(10)

where $\phi = k_x m + k_y h n - \omega t$ is the carrier wave phase. Here, the wave vector **k** is expressed as $\mathbf{k} = k_x \mathbf{e}_x + k_y \mathbf{e}_y$ and $\omega(\mathbf{k})$ corresponds to its temporal frequency. Similarly, for B sites we look for solutions of the form

$$\beta_{m,n}(t) = \rho e^{i\phi} p(x,y,\tau,T) + \rho^2 [q_0(x,y,\tau,T) + e^{i\phi} q_1(x,y,\tau,T) + e^{2i\phi} q_2(x,y,\tau,T)]$$
$$+ \rho^3 [r_0(x,y,\tau,T) + e^{i\phi} r_1(x,y,\tau,T) + e^{2i\phi} r_2(x,y,\tau,T) + e^{3i\phi} r_3(x,y,\tau,T)] + \ldots$$

(11)

It is should be mentioned that, in Eqs. (10) and (11), $f$, $g_l$, $h_l$, $p$, $q_l$ and $r_l$ ($l = 0,1,2,3$) are all functions of those slow variables, i.e., $x$, $y$, $\tau$, and $T$. If the ansatzes (10) and (11) are inserted into the equation of motion corresponding to the A and B sites, then we equate the coefficients of each harmonic frequency at each order of $\rho$ so as to obtain two sets of equations.

First, let us analyze the $O(\rho e^{i\phi})$ terms of Eqs. (7)–(8), which yields two equations on $f$ and $p$ as follows

$$\mathbf{M}\begin{pmatrix} f \\ p \end{pmatrix} := \begin{pmatrix} \omega_0 - \omega - A & -\mu \\ -\mu^* & \omega_0 - \omega + A \end{pmatrix} \begin{pmatrix} f \\ p \end{pmatrix}$$
$$= 0, \quad (12)$$

where

$$\mu = JS(e^{ik_x + ihk_y} + e^{ik_x - ihk_y} + e^{-2ik_x}), \quad (13)$$

$$A = 4DS[\cos(3k_x) - \cos(hk_y)]\sin(hk_y), \quad (14)$$

and $\mu^*$ denotes the complex conjugate of $\mu$. Here, we write $\mu = |\mu|e^{-i\theta}$, where the magnitude is given by

$$|\mu| = JS\sqrt{3 + 4\cos(3k_x)\cos(hk_y) + 2\cos(2hk_y)} \quad (15)$$

Because we focus on solutions where $(f, p)^T \neq 0$, Eq. (12) can be regarded as an eigenvalue problem, where the eigenvalue is $\omega$. Thus, the determinant of the matrix must be equal to zero, which gives the magnon dispersion relation, namely,

$$\omega = \omega_0 \pm \sqrt{|\mu|^2 + A^2}. \quad (16)$$

The minus sign in Eq. (16) corresponds to an acoustic branch, i.e., surface in $(k_x, k_y, \omega)$ space of lower frequencies, which is written as $\omega_{ac} = \omega_0 - \sqrt{|\mu|^2 + A^2}$. Furthermore, the surface for the plus sign in Eq. (16) is called the optical branch, which is denoted via $\omega_{opt} = \omega_0 + \sqrt{|\mu|^2 + A^2}$. The acoustic branch corresponds to magnon frequencies in the region $\omega_0 - 3JS \leq \omega_{ac} \leq \omega_0 - 3\sqrt{3}DS$, and meanwhile the optical branch satisfies $\omega_0 + 3\sqrt{3}DS \leq \omega_{opt} \leq \omega_0 + 3JS$. At the center of the Brillouin zone $\mathbf{k}=0$, the magnon frequency spectrum possesses a lower cutoff $\omega_{min} = \omega_0 - 3JS$ for the acoustic branch and an upper cutoff $\omega_{max} = \omega_0 + 3JS$ for the optical branch.

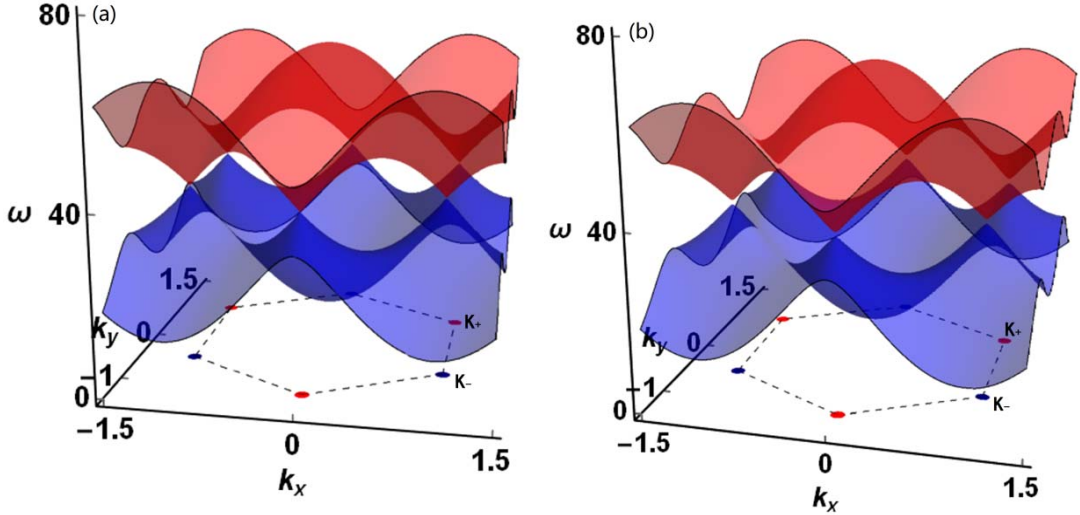

**Fig. 3.** The magnon frequency band structure for the honeycomb ferromagnet: (a) $D=0$, (b) $D=0.05$. The other parameters are as follows: $J=1$, $S=10$, $H_z=10$, $g=1$, and $\mu_B=1$. Note that the choice of these system parameters is based on Refs.[7,14,18]. In many theoretical papers, dimensionless system parameters have been often used for convenience. In this treatment, $J$, $\hbar$, $g$, and $\mu_B$ can be set to 1 meanwhile both $H_z$ and $D$ are two tunable dimensionless parameters. In addition, the setting of the spin value is based on Ref. [14]. In the rest of the paper, such dimensionless system parameters will be considered.

The complete magnon frequency dispersion relation is shown in Fig. 3. Obviously, the magnon frequency spectrum contains two branches: the acoustic "down" and optical "up" branches. It is shown that there are six Dirac points. Physically, the appearance of these Dirac points are owing to the underlying symmetries of the honeycomb lattices. Many interesting phenomena are linked with the Dirac points. Actually, there exists only two inequivalent Dirac points located at $\mathbf{K}_{\pm} = (\frac{\pi}{3}, \pm\frac{\pi}{3\sqrt{3}})$. In the absence of the DM interaction ($D=0$), the upper and the lower band meet at the Dirac points, see Fig. 3(a). After introducing the DM interaction, a band gap $\Delta\omega = 6\sqrt{3}DS$ forms at these points as depicted in Fig. 3(b). Opening of a band gap at the Dirac points is due to the breaking of the inversion symmetry of the honeycomb lattice. The appearance of the band gap makes the band structure topologically nontrivial [2]. The Chern numbers for the magnon bands are evaluated as $C_{\pm} = \pm 1$.

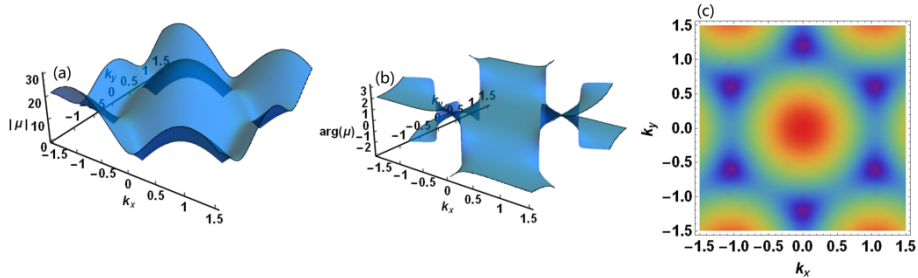

**Fig. 4.** (a) The surface diagram for $|\mu|$; (b) The surface diagram for $\arg(\mu)$; (c) The density diagram for $|\mu|$. The corresponding parameters are $J=1$ and $S=10$.

Fig. 3 exhibits the hexagonal symmetry of the honeycomb ferromagnetic lattice. Compared Eq. (15) with (16), we deduce that $\mu = 0$ appears at the Dirac points. Fig. 4 displays the magnitude and argument of $\mu$ as function of $(k_x, k_y)$. In Fig. 4, one can see clearly that $\mu$ is equal to zero at the Dirac points. Furthermore, the hexagonal symmetry can also be clearly observed in Fig. 4(c).

It is not sufficient to get the eigenvalue, $\omega$, one also need to obtain the solutions for $f$ and $p$, therefore we require the eigenvector of Eq. (12) as well. On account of $\det(\mathbf{M}) = 0$, solutions can be set to $(f, p)^T = f(1, C)^T$. By solving Eq. (12), one obtains $C(k_x, k_y)$ for $\omega_{ac}$ and $\omega_{opt}$, namely,

$$C_{ac} = \frac{\sqrt{|\mu|^2 + A^2} - A}{|\mu|} e^{i\theta}, \quad C_{opt} = -\frac{\sqrt{|\mu|^2 + A^2} + A}{|\mu|} e^{i\theta}. \tag{17}$$

At $O(\rho^2 e^{0i\phi})$, one has

$$\omega_0 g_0 - 3JSq_0 = 0, \quad \omega_0 q_0 - 3JSg_0 = 0. \tag{18}$$

By solving the above equation, it is easy to obtain $g_0 = q_0 = 0$.

Considering the terms at $O(\rho^2 e^{2i\phi})$, we can get

$$g_2 = q_2 = 0 \tag{19}$$

According to our asymptotic scheme of substituting Eqs. (10) and (11) into the equations of motion (7) and (8) and equating terms of $O(\rho^2 e^{i\phi})$, we obtain

$$\mathbf{M}\begin{pmatrix} g_1 \\ q_1 \end{pmatrix} = \begin{pmatrix} if_\tau + i\mu_1 p_x + i\mu_2 p_y + id_1 f_x + id_2 f_y \\ ip_\tau - id_1 p_x - id_2 p_y + i\mu_1^* f_x + i\mu_2^* f_y \end{pmatrix} \tag{20}$$

with

$$\mu_1 = iJS(2e^{-2ik_x} - e^{ik_x - ihk_y} - e^{ik_x + ihk_y}), \quad \mu_2 = ihJS(e^{ik_x - ihk_y} - e^{ik_x + ihk_y})$$

$$d_1 = 12DS \sin(3k_x) \sin(hk_y), \quad d_2 = -4DSh[\cos(3k_x)\cos(hk_y) - \cos(2hk_y)]$$

$$\tag{21}$$

where $\mathbf{M}$ is the matrix given in Eq. (12).

Since $\det(\mathbf{M}) = 0$, an equation such as Eq. (20), which can be recast into $\mathbf{M}(g_1, q_1)^T = \mathbf{F}$, either does not exist solutions, or has a whole family of solutions for $(g_1, q_1)^T$. In the light of the Fredholm alternative [27], the existence of solutions depends on $\mathbf{F}$. Solutions exist only when the RHS of Eq. (20), i.e. $\mathbf{F}$, is in the range of the matrix $\mathbf{M}$. From Ref. [28], calculating the eigenvector of the matrix $\mathbf{M}$ can yield

$$Range_{ac} = K\begin{pmatrix} -(\sqrt{|\mu|^2+A^2}-A)e^{-i\theta} \\ |\mu| \end{pmatrix} = K\begin{pmatrix} -\frac{(\sqrt{|\mu|^2+A^2}-A)}{|\mu|}e^{-i\theta} \\ 1 \end{pmatrix},$$

$$Range_{opt} = K\begin{pmatrix} (\sqrt{|\mu|^2+A^2}+A)e^{-i\theta} \\ |\mu| \end{pmatrix} = K\begin{pmatrix} \frac{(\sqrt{|\mu|^2+A^2}+A)}{|\mu|}e^{-i\theta} \\ 1 \end{pmatrix}.$$

(22)

Because normals to these directions have the following forms

$$\mathbf{n}_{ac} = \begin{pmatrix} -C_{opt} \\ 1 \end{pmatrix} = \begin{pmatrix} \frac{1}{C^*_{ac}} \\ 1 \end{pmatrix}, \quad \mathbf{n}_{opt} = \begin{pmatrix} -C_{ac} \\ 1 \end{pmatrix} = \begin{pmatrix} \frac{1}{C^*_{opt}} \\ 1 \end{pmatrix}, \quad (23)$$

the circumstance that $\mathbf{F} \in$ Range means $\mathbf{n} \cdot \mathbf{F} = 0$. It should be pointed out that in both the optical and the acoustic branches, Eq. (23) suggests $\mathbf{n} = (1/C^*, 1)^T$.

Let us also come back that the leading order quantities, $p$ and $f$ satisfy the relation $p = Cf$, where both $C$ and $\mathbf{n}$ have different forms for the acoustic and optical branches, given by Eq. (17). Taking advantage of $p = Cf$ and $\mathbf{n} \cdot \mathbf{F} = 0$, we can get the following equation

$$f_\tau + \frac{[(1-|C|^2)d_1 + C\mu_1 + C^*\mu_1^*]}{(|C|^2+1)}f_x + \frac{[C\mu_2 + C^*\mu_2^* + (1-|C|^2)d_2]}{(|C|^2+1)}f_y = 0, \quad (24)$$

which suggests that $f$ (and therefore $p$ as well) is an travelling wave solution. Thus, we look for the travelling wave solution of the form

$$f(x,y,\tau,T) \equiv f(z,w,T), \quad p(x,y,\tau,T) \equiv p(z,w,T), \quad (25)$$

where $z = x - u\tau$ and $w = y - v\tau$ are new position coordinates in the horizontal and vertical directions, respectively. It is not hard to derive the $x$ and $y$ components of the group velocity:

$$u = \partial\omega/\partial k_x = \frac{C\mu_1 + C^*\mu_1^* + (1-|C|^2)d_1}{(|C|^2+1)}, \quad (26)$$

$$v = \partial\omega/\partial k_y = \frac{C\mu_2 + C^*\mu_2^* + (1-|C|^2)d_2}{(|C|^2+1)}. \quad (27)$$

The magnitude of the group velocity ( or the overall speed) can be written as

$$c = \sqrt{u^2+v^2}. \quad (28)$$

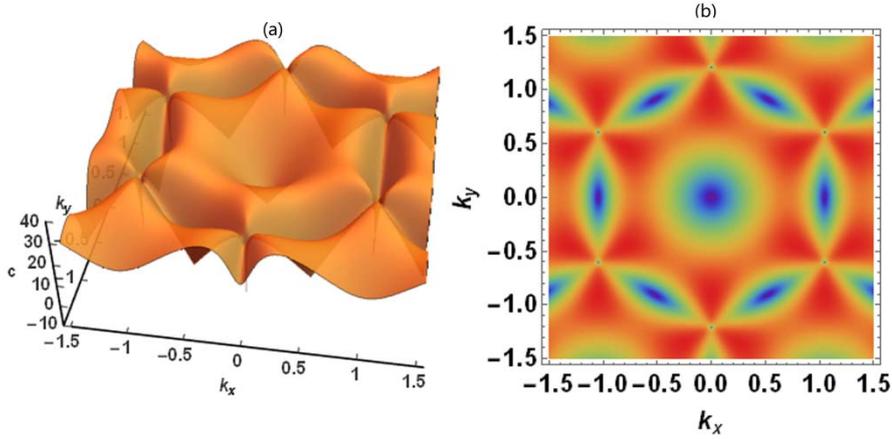

**Fig. 5.** (a) The surface diagram for $c$; (b) The density diagram for $c$. The relevant parameters are $J=1$, $S=10$, $D=0.01$, $H_z=0$, $g=1$, and $\mu_B=1$.

Considering that the expressions for $\omega_{ac}$ and $\omega_{opt}$ are distinct, hence Eqs. (26)and(27) also produces different formulas for $u_{ac}$ and $u_{opt}$ (similarly for $v_{ac}$ and $v_{opt}$). From Eqs.(16), (26) and (27), however, we conclude $u_{ac} = -u_{opt}$ and $v_{ac} = -v_{opt}$ for the same wavenumber. Thus, the expressions for $c_{ac}$ and $c_{opt}$ are same so that we can write down $c_{ac} = c_{opt} = c$ for any given wavenumber $\mathbf{k} = k_x\mathbf{e_x} + k_y\mathbf{e_y}$. Fig. 5 shows the magnitude of the group velocity as functions of the wavenumbers. Note that at the Dirac points, where $|\mu|=0$, $u_{ac}$, $u_{opt}$, $v_{ac}$, and $v_{opt}$ are equal to zero. Therefore, the magnitude of the group velocity is equal to zero at the Dirac points, as displayed in Fig. 5(b).

In the previous calculation, the existence condition on solutions for the RHS of Eq. (20) has been obtained. However, we still do not know the quantities $g_1$ and $q_1$. Since the solution to Eq. (20) is in fact degenerate, the corresponding one-parameter family of solutions may has the following form

$$\begin{pmatrix} g_1 \\ q_1 \end{pmatrix} = \bar{g}_1 \begin{pmatrix} 1 \\ C \end{pmatrix} + \hat{g}_1 \begin{pmatrix} 1 \\ 0 \end{pmatrix}, \tag{29}$$

where $\hat{g}_1$ is determined via Eq. (20), and $\bar{g}_1$ is an arbitrary quantity. Note that the two equations for $\hat{g}_1$ generated by Eq.(20) are identical. Thus, we can obtain

$$\hat{g}_1 = \frac{is}{2\sqrt{|\mu|^2 + A^2}}[(2d_1 + C\mu_1 - \frac{\mu_1^*}{C})f_z + (2d_2 + C\mu_2 - \frac{\mu_2^*}{C})f_w], \quad s = \begin{cases} + & \text{acoustic} \\ - & \text{optical} \end{cases} \tag{30}$$

The above equation can be rewritten into $\hat{g}_1 = i\hat{u}f_z + i\hat{v}f_w$, where $\hat{u}$ and $\hat{v}$ are respectively given by

$$\hat{u} = \frac{s}{2\sqrt{|\mu|^2 + A^2}}(2d_1 + C\mu_1 - \frac{\mu_1^*}{C}), \quad \hat{v} = \frac{s}{2\sqrt{|\mu|^2 + A^2}}(2d_2 + C\mu_2 - \frac{\mu_2^*}{C}). \tag{31}$$

At $O(\rho^3 e^{0i\phi})$, one can get the following equation

$$\omega_0 h_0 - 3JSr_0 = 0, \quad \omega_0 r_0 - 3JSh_0 = 0. \tag{32}$$

By solving the above equation, we have $h_0 = r_0 = 0$.

When Eqs. (10) and (11) are substituted into the equations of motion (7) and (8), and terms of $O(\rho^3 e^{i\phi})$ are equated, we can obtain the final equation, which reads

$$\mathbf{M}\begin{pmatrix} h_1 \\ r_1 \end{pmatrix} = \begin{pmatrix} C_1 \\ C_2 \end{pmatrix}. \tag{33}$$

Here, the matrix $\mathbf{M}$ has been defined via Eq.(12), and the RHS components have the following forms

$$C_1 = if_T + ig_{1\tau} + id_1 g_{1x} + id_2 g_{1y} + i\mu_1 q_{1x} + i\mu_2 q_{1y} + d_3 f_{xx} + d_4 f_{xy} + d_5 f_{yy} + \mu_3 p_{yy} + i\mu_2 p_{xy} + \mu_4 p_{xx}$$
$$- \frac{\mu^*}{4S} f^2 p^* + 3Jf|p|^2 - \frac{\mu}{4S}|p|^2 p - d_6 |f|^2 f,$$

$$C_2 = ip_T + iq_{1\tau} + i\mu_1^* g_{1x} + i\mu_2^* g_{1y} - id_2 q_{1y} - id_1 q_{1x} + \mu_3^* f_{yy} - i\mu_2^* f_{xy} + \mu_4^* f_{xx} - d_3 p_{xx} - d_4 p_{xy}$$
$$- d_5 p_{yy} - \frac{\mu}{4S} p^2 f^* + 3Jp|f|^2 - \frac{\mu^*}{4S}|f|^2 f + d_6 |p|^2 p$$

(34)

with

$$\mu_3 = h^2 JSe^{ik_x}\cos(hk_y), \qquad \mu_4 = JS(\frac{1}{2}e^{ik_x - ihk_y} + \frac{1}{2}e^{ik_x + ihk_y} + 2e^{-2ik_x}),$$
$$d_3 = 18DS\cos(3k_x)\sin(hk_y), \qquad d_4 = 12DhS\sin(3k_x)\cos(hk_y),$$
$$d_5 = 2Dh^2 S[\cos(3k_x) - 4\cos(hk_y)]\sin(hk_y), \quad d_6 = 2D[\cos(3k_x) - \cos(hk_y)]\sin(hk_y).$$

(35)

As in the case of the equations at $O(\rho^2 e^{i\phi})$, to ensure that the present equations have solutions, the consistency condition $\mathbf{n} \cdot \begin{pmatrix} C_1 \\ C_2 \end{pmatrix} = 0$ must be met. There is no need to find $h_1$ and $r_1$. Taking $g_1, q_1$ as given via Eq. (29) with $\bar{g}_1 = -\hat{g}_1/(1+|C|^2)$ and making use of $p = Cf$, one can get a single partial differential equation on $f$, which has the following form

$$f_T + P_1 f_{zz} + P_2 f_{zw} + P_3 f_{ww} + Q|f|^2 f = 0 \qquad (36)$$

with

$$P_1 = -\frac{|C|^2 C^* \mu_1^* - C\mu_1 + 2d_1 |C|^2}{(1+|C|^2)^2}\hat{u} + d_3 \frac{1-|C|^2}{1+|C|^2} + \frac{C\mu_4 + C^* \mu_4^*}{1+|C|^2},$$

$$P_2 = \frac{1-|C|^2}{1+|C|^2}d_4 + i\frac{\mu_2 C - C^* \mu_2^*}{1+|C|^2} + \frac{C\mu_1 - |C|^2 C^* \mu_1^*}{(1+|C|^2)^2}\hat{v}$$
$$- 2\frac{|C|^2 (d_1 \hat{v} + d_2 \hat{u})}{(1+|C|^2)^2} + \frac{C\mu_2 - |C|^2 C^* \mu_2^*}{(1+|C|^2)^2}\hat{u},$$

$$P_3 = -\frac{|C|^2 C^*}{(1+|C|^2)^2}\mu_2^*\hat{v} + \frac{C}{(1+|C|^2)^2}\mu_2\hat{v} - 2d_2\frac{|C|^2}{(1+|C|^2)^2}\hat{v}$$

$$+\frac{1-|C|^2}{1+|C|^2}d_5 + \frac{C\mu_3 + C^*\mu_3^*}{1+|C|^2},$$

$$Q = (|C|^2-1)d_6 - \frac{C^*}{1+|C|^2}\frac{\mu^*}{2S} - \frac{C|C|^2}{1+|C|^2}\frac{\mu}{2S} + 6\frac{|C|^2}{1+|C|^2}J.$$

(37)

It is obvious that Eq. (36) is a 2D nonlinear Schrödinger (NLS) equation. With the exception of **k** at the Dirac points, $P_1$, $P_2$, and $P_3$ also satisfy the following relationship

$$P_1 = \frac{1}{2}\frac{\partial^2\omega}{\partial k_x^2},\quad P_2 = \frac{\partial^2\omega}{\partial k_x \partial k_y},\quad P_3 = \frac{1}{2}\frac{\partial^2\omega}{\partial k_y^2}, \tag{38}$$

which has been verified by the use of Mathematica. By means of the Hirota bilinear method [29], the bright and dark (one) soliton solutions of Eq. (36) can be easily obtained.

To construct bright soliton solution of the 2D NLS equation (36), we employ the following variable transformation

$$f(T,z,w) = \frac{G(T,z,w)}{F(T,z,w)}, \tag{39}$$

where $G(T,z,w)$ is a complex differentiable function and $F(T,z,w)$ is a real differentiable function with respect to $T$, $z$, and $w$. With Eq. (39), the 2D NLS equation (36) can be recast into the following bilinear equation

$$[iD_T + P_1 D_z^2 + P_3 D_w^2 + P_2 D_z D_w](G\cdot F) = 0,$$

$$[P_1 D_z^2 + P_3 D_w^2 + P_2 D_z D_w](F\cdot F) - Q|G|^2 = 0,$$

(40)

where $D_T$, $D_z$, and $D_w$ are the Hirota bilinear operators.

Then, in order to get the bright soliton solution, we need to consider the following assumption

$$G(T,z,w) = \varepsilon G_1(T,z,w), \quad F(T,z,w) = 1 + \varepsilon^2 F_2(T,z,w) \tag{41}$$

Substituting Eq. (41) into the bilinear forms (40) and collecting the terms with the same power of $\varepsilon$, one can obtain the following bright soliton solution:

$$f(T,z,w) = \frac{e^{a_1 z + a_2 w}}{1 + \xi e^{2a_1 z + 2a_2 w}} e^{isT} \tag{42}$$

with

$$s = P_1 a_1^2 + P_2 a_1 a_2 + P_3 a_2^2, \quad \xi = \frac{Q}{8(P_1 a_1^2 + P_2 a_1 a_2 + P_3 a_2^2)}, \tag{43}$$

where both $a_1$ and $a_2$ are the real constants. Without loss of generality, $\varepsilon$ has been set to $\varepsilon = 1$ in Eq. (42).

Next, to construct dark soliton solution of the 2D NLS equation (36), the bilinear equations (40) should be changed into the following form

$$[iD_t + P_1 D_z^2 + P_3 D_w^2 + P_2 D_z D_w](G \cdot F) - \lambda GF = 0,$$

$$[P_1 D_z^2 + P_3 D_w^2 + P_2 D_z D_w](F \cdot F) - \lambda F^2 - D|G|^2 = 0. \tag{44}$$

Here, $\lambda$ is a constant, which can be confirmed in the following calculation. Mathematically, to obtain the dark soliton solution, one may assume the following expressions

$$G(T,z,w) = G_0(T,z,w)[1 + \varepsilon G_1(T,z,w)], \quad F(T,z,w) = 1 + \varepsilon F_1(T,z,w). \tag{45}$$

By solving the bilinear equation (44) with the substitution of assumption (45), we can get the dark soliton solution, which is as follows

$$f(T,z,w) = \eta \tanh[\frac{-P_2 b_2 + \sqrt{(P_2^2 - 4P_1 P_3)b_2^2 - 8P_1 Q \eta^2}}{4P_1} z + \frac{b_2 w}{2}] e^{i\eta^2 QT}, \qquad (46)$$

where $\eta$ and $b_2$ are two real constants.

Let us consider a specific 2D NLS equation, which has the following form

$$f_T + P(f_{zz} + f_{ww}) + Q|f|^2 f = 0. \qquad (47)$$

In general, there exists two different cases for Eq. (47), i.e., the anomalous dispersion regime where $sign(PQ) > 0$, also called the focusing 2D NLS equation; and the normal dispersion regime where $sign(PQ) < 0$, also referred to as the defocusing 2D NLS equation. Mathematically, the focusing 2D NLS equation supports bright soliton solutions while the defocusing 2D NLS equation exists dark soliton solutions[30]. In the next section, this fact will be considered for the analysis of soliton solutions.

## 4. Nonlinear localized excitations

From Eq. (36), one can see clearly that the envelope function $f$ obeys the 2D NLS equation. Its bright and dark solition solutions have been constructed exactly by using the Hirota bilinear method. In this section, we shall apply these solition solutions to obtain the analytical forms of nonlinear localized excitations for two types of special wave vectors: (i) $\mathbf{k} = 0$, (ii) $\mathbf{k}$ near the Dirac points.

### A. The Brillouin zone center modes

In general, nonlinear localized modes at the Brillouin zone center(i.e., $\mathbf{k} = 0$) can be known as the Brillouin zone center modes. Here, considering that the magnon frequency spectrum of the present model contains two branches, therefore we need to analyze the acoustic mode and optical mode at the Brillouin zone center, respectively.

For the acoustic mode at the Brillouin zone center, we have $P_{1,ac} = P_{3,ac} = 3JS$, $P_{2,ac} = 0$, and $Q_{ac} = 3J/2$. What is more, it is noted that $\omega_{ac} = \omega_{min} = \omega_0 - 3JS$, $u_{ac} = v_{ac} = 0$, and $C_{ac} = 1$. In this case, Eq. (36) can be reduced as a focusing 2D NLS equation, which supports the bright soliton solution. Thus, with the bright soliton solution (42), it is not difficult to construct the analytical form of the acoustic mode at the Brillouin zone center, namely,

$$\alpha_{m,n}(t) = \beta_{m,n}(t) \approx \rho e^{-i\Omega_1 t} \frac{e^{\rho a_1 m + \rho a_2 h n}}{1 + \frac{1}{16S(a_1^2 + a_2^2)} e^{2\rho a_1 m + 2\rho a_2 h n}}, \qquad (48)$$

where $\Omega_1 = \omega_{min} - 3\rho^2(a_1^2 + a_2^2)JS$. It is obvious that Eq. (48) is a 2D discrete breather with a bright localized structure, whose eigenfrequency $\Omega_1$ is below the bottom of magnon acoustic frequency band. Fig. 6 shows the square of the coherent-state amplitude $|\alpha_{m,n}|^2$ (the same plot for $|\beta_{m,n}|^2$) for the 2D discrete breather, whose the spatially localized structure can stay the same.

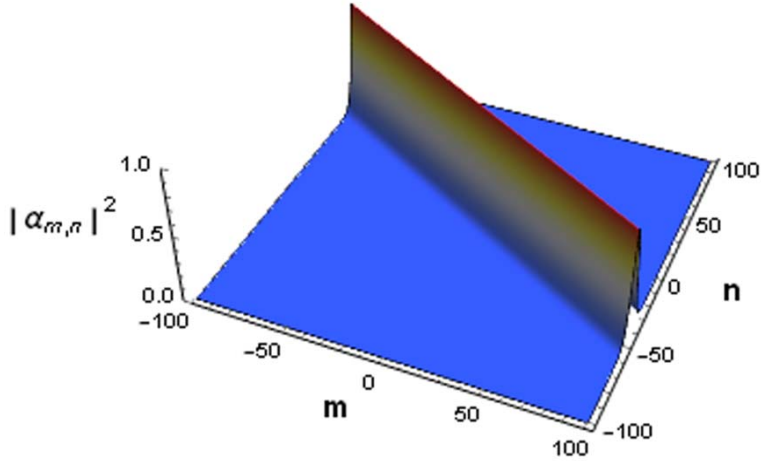

**Fig. 6**. Plot of the 2D bright discrete breather via Eq. (48). The relevant parameters are $J=1$, $S=10$, $H_z=10$, $g=1$, $\mu_B=1, a_1=1, a_2=1$, and $\rho=0.1$. For all $t$, the plots are the same.

Let us turn our attention to the optical mode at the Brillouin zone center. Then, we have $P_{1,opt} = P_{3,opt} = -3JS$, $P_{2,opt} = 0$, and $Q_{opt} = 9J/2$. In this case, Eq. (36) is a defocusing 2D NLS equation, which exists the dark soliton solution. Furthermore, we note that $\omega_{ac} = \omega_{max} = \omega_0 + 3JS$, $u_{ac} = v_{ac} = 0$, and $C_{ac} = -1$. Thus, we can construct the analytical form of the optical mode at the Brillouin zone center:

$$\alpha_{m,n}(t) \approx \rho\eta e^{-i\Omega_2 t} \tanh(\frac{\sqrt{3S\eta^2 - S^2 b_2^2}}{2S}\rho m + \frac{b_2 h \rho n}{2}), \qquad (49\text{ a})$$

$$\beta_{m,n}(t) \approx -\rho\eta e^{-i\Omega_2 t} \tanh(\frac{\sqrt{3S\eta^2 - S^2 b_2^2}}{2S}\rho m + \frac{b_2 h \rho n}{2}), \qquad (49\text{b})$$

where $\Omega_2 = \omega_{max} - 9J\rho^2\eta^2/2$. Eq. (49) is a 2D discrete breather with a dark localized structure, as shown in Fig. 7. The vibration frequency of the dark discrete breather is within the magnon optical frequency band, hence it is a resonant mode. Physically, the resonant dark discrete breather mode is in resonance with the magnon frequency spectrum so that it has the finite lifetime in real ferromagnetic materials[31].

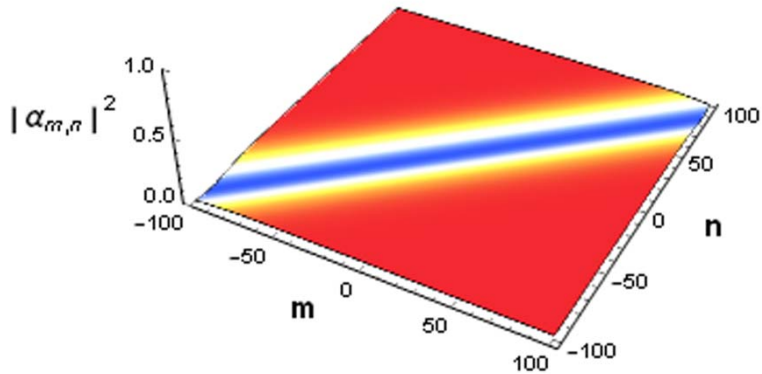

**Fig. 7**. (Color online) Plot of the 2D dark discrete breather via Eq. (49 a). The relevant parameters are $J = 1$, $S = 10$, $H_z = 10$, $g = 1$, $\mu_B = 1$, $\eta = 2$, $b_2 = 0.6$, and $\rho = 0.1$. Note that the plot for $\left|\beta_{m,n}\right|^2$ is the same with this figure. Since the localized structure of the 2D dark discrete breather remains unchanged, the plots are the same for all $t$.

### B. Nonlinear excitations in the vicinity of the Dirac points

In our work, the studies on the nonlinear excitations of the system is based on the amplitude equation (36). At the Dirac points, however, $P_1$, $P_2$ and $P_3$ lose their physical meaning as that this amplitude equation is invalid at these points. Furthermore, $Q$ is continuously differentiable at the Dirac points only if $|C|$ is finite. Based on this fact, we focus on nonlinear excitations near Dirac points(i.e., the wavenumber of the traveling wave satisfying $|\mathbf{k} - \mathbf{K}_\pm| \ll 1$) in the following passages.

Here, we first analyze nonlinear excitations in the vicinity of $\mathbf{K}_+ = (\frac{\pi}{3}, \frac{\pi}{3\sqrt{3}})$. Since it is impossible to directly confirm the type of soliton solutions to Eq. (36) by means of Eq.(37), we shall turn to adopt an approximate method to obtain the analytical forms of these nonlinear excitations. Considering that $C_{opt}$ is finite but $C_{ac}$ rounds towards infinity when $\mathbf{k} \to \mathbf{K}_+$, hence we only study the optical mode near $\mathbf{K}_+$. Applying the Taylor expansion for $\omega_{opt}$ and keeping second order, then we have

$$\omega_{opt} = \omega_0 + 3\sqrt{3}DS + \frac{\sqrt{3}(J^2 - 18D^2)S}{2D}(k_x - \frac{\pi}{3})^2 + \frac{\sqrt{3}(J^2 - 18D^2)S}{2D}(k_y - \frac{\pi}{3\sqrt{3}})^2. \qquad (50)$$

Due to $\omega_{opt} > \omega_0 + 3\sqrt{3}DS$, the system parameters must satisfy $0 < D < J/(3\sqrt{2})$. By substituting Eq. (50) into Eqs. (26) and (27), it is easy to derive the group velocity of spin wave in the vicinity of $\mathbf{K}_+$, which is

$$u_{opt} \approx \frac{\sqrt{3}(J^2 - 18D^2)S}{D}(k_x - \frac{\pi}{3}), \quad v_{opt} \approx \frac{\sqrt{3}(J^2 - 18D^2)S}{D}(k_y - \frac{\pi}{3\sqrt{3}}). \quad (51)$$

Moreover, approximate expressions for $P_{1,opt}$, $P_{2,opt}$, and $P_{3,opt}$ can also be gotten via substitution Eq. (50) into Eq. (38). In order to proceed, we only consider the lowest order approximation for $Q_{opt}$ because it is almost flat near $\mathbf{K}_+$. Then, Eq. (36) can be reduced as

$$f_T + \frac{\sqrt{3}(J^2 - 18D^2)S}{2D}(f_{zz} + f_{ww}) + \frac{3\sqrt{3}}{2}D|f|^2 f = 0. \quad (52)$$

Here, we must consider the confine condition $0 < D < J/(3\sqrt{2})$, then Eq. (52) is a focusing 2D NLS equation, which exists the bright soliton solution. With Eq. (42), we can obtain the bright soliton solutions for $\alpha_{mn}$ and $\beta_{mn}$, which are respectively given by

$$\alpha_{m,n}(t) \approx \rho e^{i(k_x m + k_y h n) - \Omega_3 t} \frac{e^{\rho a_1 [m - \frac{\sqrt{3}(J^2-18D^2)S}{D}(k_x - \frac{\pi}{3})t] + \rho a_2 (hn - \frac{\sqrt{3}(J^2-18D^2)S}{D}[k_y - \frac{\pi}{3\sqrt{3}})t]}}{1 + \frac{3D^2}{8(J^2 - 18D^2)S(a_1^2 + a_2^2)} e^{2\rho a_1 [m - \frac{\sqrt{3}(J^2-18D^2)S}{D}(k_x - \frac{\pi}{3})t] + 2\rho a_2 [hn - \frac{\sqrt{3}(J^2-18D^2)S}{D}(k_y - \frac{\pi}{3\sqrt{3}})t]}}, \quad (53a)$$

$$\beta_{m,n}(t) \approx C_{opt} \rho e^{i(k_x m + k_y h n) - \Omega_3 t} \frac{e^{\rho a_1 [m - \frac{\sqrt{3}(J^2-18D^2)S}{D}(k_x - \frac{\pi}{3})t] + \rho a_2 (hn - \frac{\sqrt{3}(J^2-18D^2)S}{D}[k_y - \frac{\pi}{3\sqrt{3}})t]}}{1 + \frac{3D^2}{8(J^2 - 18D^2)S(a_1^2 + a_2^2)} e^{2\rho a_1 [m - \frac{\sqrt{3}(J^2-18D^2)S}{D}(k_x - \frac{\pi}{3})t] + 2\rho a_2 [hn - \frac{\sqrt{3}(J^2-18D^2)S}{D}(k_y - \frac{\pi}{3\sqrt{3}})t]}}, \quad (53b)$$

where $\Omega_3 = \omega_{opt} - \rho^2 \frac{\sqrt{3}(J^2 - 18D^2)S}{2D}(a_1^2 + a_2^2)$ is the vibration frequency of the this bright soliton. In the present approximation, the bright soliton (53) can be viewed as a bulk gap solition when soliton parameters meet the condition: $a_1 \rho > |k_x - \pi/3|$, $a_2 \rho > |k_y - \pi/(3\sqrt{3})|$. Fig. 8 shows the waveform of this bulk gap soliton at different times. It is obvious that it propagates along the $\mathbf{e}_y$ direction, i.e., the positive $n$-axis. From Eq. (53), one can deduce that the gap soliton velocity is equal to the group velocity of optical spin waves in the vicinity of $\mathbf{K}_+$. So, its propagation direction depends on the wave number **k**. Thus, we can understand the propagation direction of the gap soliton in Fig. 8.

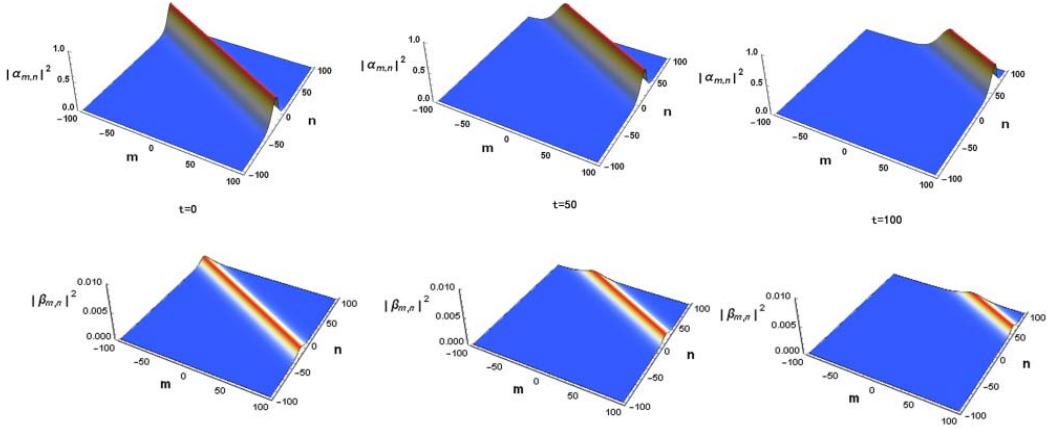

**Fig. 8**. Plot of the bulk gap solition via Eq. (53). The relevant parameters are $J=1$, $S=10$, $H_z=10$, $g=1$, $\mu_B=1$, $a_1=1$, $b_1=1$, $D=0.2$ and $\rho=0.05$. The wavenumber is fixed as $k_x = \dfrac{\pi}{3}$, $k_y = \dfrac{\pi}{3\sqrt{3}} + \dfrac{\pi}{100}$.

Next, let turn our attention to nonlinear excitations near $\mathbf{K}_- = (\dfrac{\pi}{3}, -\dfrac{\pi}{3\sqrt{3}})$. Since $C_{ac}$ is finite but $C_{opt}$ tends to infinity for the case of $\mathbf{k} \to \mathbf{K}_-$, we only need to consider the acoustic mode near $\mathbf{K}_-$. Similarly, we use the Taylor expansion for $\omega_{ac}$ and keep the second order, which yields

$$\omega_{ac} = \omega_0 - 3\sqrt{3}DS - \dfrac{\sqrt{3}(J^2 - 18D^2)S}{2D}(k_x - \dfrac{\pi}{3})^2 - \dfrac{\sqrt{3}(J^2 - 18D^2)S}{2D}(k_y + \dfrac{\pi}{3\sqrt{3}})^2. \quad (54)$$

Due to $\omega_{ac} < \omega_0 - 3\sqrt{3}DS$, the system parameters also have to meet $0 < D < J/(3\sqrt{2})$. With Eq. (54), we can derive the group velocity of spin waves in the vicinity of $\mathbf{K}_-$, which is

$$u_{ac} = -\frac{\sqrt{3}(J^2-18D^2)S}{D}(k_x - \frac{\pi}{3}), \quad v_{ac} = -\frac{\sqrt{3}(J^2-18D^2)S}{D}(k_y + \frac{\pi}{3\sqrt{3}}). \tag{55}$$

What is more, by inserting Eq. (54) into Eq. (38), approximate analytical forms of $P_{1,ac}$, $P_{2,ac}$, and $P_{3,ac}$ can be obtained. As the previous analysis, the lowest order approximation for $Q_{opt}$ is considered. Thus, the amplitude equation for the acoustic mode near $\mathbf{K}_-$ can be approximately written as

$$f_T - \frac{\sqrt{3}(J^2-18D^2)S}{2D}(f_{zz} + f_{ww}) - \frac{3\sqrt{3}}{2}D|f|^2 f = 0. \tag{56}$$

Under the restriction $0 < D < J/(3\sqrt{2})$, Eq. (56) is also a focusing 2D NLS equation, which also supports the bright soliton solution. Similarly, we can use Eq. (42) to construct the bright soliton solutions for $\alpha_{mn}$ and $\beta_{mn}$. These bright soliton solutions read

$$\alpha_{m,n}(t) \approx \rho e^{i(k_x m + k_y h n) - \Omega_{ac} t} \frac{e^{\rho a_1 [m - \frac{\sqrt{3}(J^2-18D^2)S}{D}(k_x - \frac{\pi}{3})t] + \rho a_2 (hn - \frac{\sqrt{3}(J^2-18D^2)S}{D}[k_y + \frac{\pi}{3\sqrt{3}}]t)}}{1 + \frac{3D^2}{8(J^2-18D^2)S(a_1^2+a_2^2)} e^{2\rho a_1 [m + \frac{\sqrt{3}(J^2-18D^2)S}{D}(k_x - \frac{\pi}{3})t] + 2\rho a_2 [hn + \frac{\sqrt{3}(J^2-18D^2)S}{D}(k_y + \frac{\pi}{3\sqrt{3}})t]}}, \tag{57a}$$

$$\beta_{m,n}(t) \approx C_{ac}\rho \frac{e^{\rho a_1 [m + \frac{\sqrt{3}(J^2-18D^2)S}{D}(k_x - \frac{\pi}{3})t] + \rho a_2 (hn + \frac{\sqrt{3}(J^2-18D^2)S}{D}[k_y + \frac{\pi}{3\sqrt{3}}]t)}}{1 + \frac{3D^2}{8(J^2-18D^2)S(a_1^2+a_2^2)} e^{2\rho a_1 [m + \frac{\sqrt{3}(J^2-18D^2)S}{D}(k_x - \frac{\pi}{3})t] + 2\rho a_2 [hn + \frac{\sqrt{3}(J^2-18D^2)S}{D}(k_y + \frac{\pi}{3\sqrt{3}})t]}} e^{i(k_x m + k_y h n) - \Omega_{opt} t}, \tag{57b}$$

where $\Omega_{ac} = \omega_{ac} + \rho^2 \frac{\sqrt{3}(J^2-18D^2)S}{2D}(a_1^2 + a_2^2)$ is the vibration frequency of this bright soliton. Only if soliton parameters meet $a_1 \rho > |k_x - \pi/3|$ and $a_2 \rho > |k_y + \pi/(3\sqrt{3})|$, then the bright soliton (57) may be a bulk gap solition. Fig. 9 displays the shapes of the bulk gap soliton (57) at $t=0$, $t=50$, and $t=100$. It is clearly seen that this gap soliton propagates along the $-\mathbf{e}_y$ direction, i.e., the negative $n$-axis. From Eq. (57), we can conclude that the gap soliton velocity is equal to the group velocity of acoustic spin waves near $\mathbf{K}_-$. Hence, its propagation direction is also connected with the wave number **k**. Under this consideration, it is not difficult to understand why our bulk gap soliton propagates along the negative $n$-axis in Fig. 9.

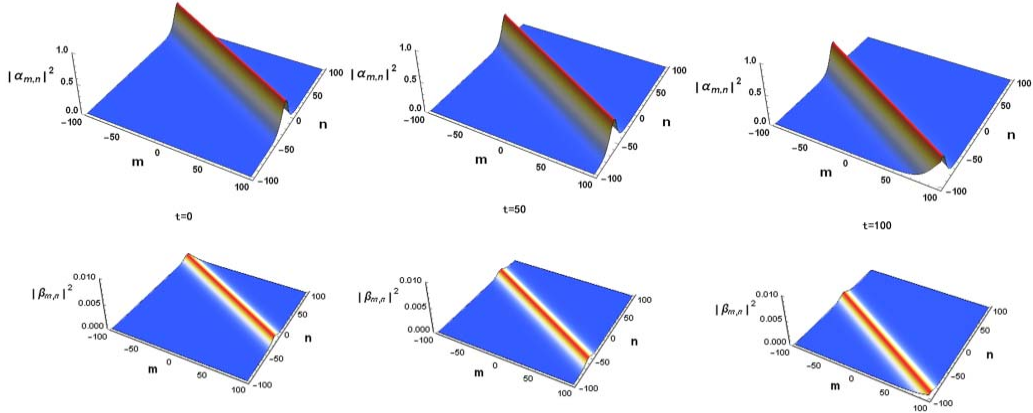

**Fig. 9**. Plot of the bulk gap solition via Eq. (57). The parameters are set to $J = 1$, $S = 10$, $H_z = 10$, $g = 1$, $\mu_B = 1$, $a_1 = 1$, $b_1 = 1$, $D = 0.2$ and $\rho = 0.05$. The wavenumber is fixed as $k_x = \frac{\pi}{3}$, $k_y = -\frac{\pi}{3\sqrt{3}} + \frac{\pi}{100}$.

## 5. Summary

To summarize, we have studied nonlinear localized excitations in a ferromagnetic honeycomb lattice with the topological band structure by using a semiclassical treatment. The 2D NLS equation was derived for the envelope function of the coherent state amplitude with the help of the semidiscrete multiple-scale method. Different types of nonlinear localized excitations can be directly obtained through similar ways. Starting from the nonlinear amplitude equations given in Eq. (36), the existence conditions and properties for the Brillouin zone center modes and nonlinear excitations near the Dirac points were analyzed, respectively. We found that, at the Brillouin zone center, a 2D bright discrete breather appears below the bottom of the acoustic branch and a resonant 2D dark discrete breather appears within the optical branch. Our results showed that, under certain conditions, bulk gap solitons are possible for the optical or acoustic mode near the Dirac points. Physically, opening of a band gap at the Dirac points requires the breaking of inversion symmetry of the honeycomb lattice. Hence, the next-nearest neighbor DM interaction plays a key role in the existence of these bulk gap solitons. Finally, we also expect that more experimental works will be devoted to observing discrete breathers and bulk gap solitons in the two dimensional honeycomb ferromagnet with the topological band structure.

**Acknowledgments**
This study was supported by the National Natural Science Foundation of China under Grant Nos. 12064011, 11875126, and 11964011, the Natural Science Fund Project of Hunan Province under Grant Nos. 2020JJ4498 and 2020JJ5453, and the Graduate Research Innovation Foundation of Jishou University under Grant No. JGY202029.